%
\magnification \magstep1
\hsize=15.4truecm
\vsize=21.8truecm
\voffset=0.9truecm
\hoffset=-0.1truecm
\normalbaselineskip=5.25mm
\baselineskip=5.25mm
\parskip=5pt
\parindent=20pt
\nopagenumbers
\headline={\ifnum\pageno>1 {\hss\tenrm-\ \folio\ -\hss} \else {\hfill}\fi}
\def\em{\sl}
%
%
\newcount\EQNcount \EQNcount=1
\newcount\CLAIMcount \CLAIMcount=1
\newcount\SECTIONcount \SECTIONcount=0
\newcount\SUBSECTIONcount \SUBSECTIONcount=0
\def\actualnumber{\number\SECTIONcount}
\newcount\timecount
\def\TODAY{\number\day~\ifcase\month\or January \or February \or March \or
  April \or May \or June
  \or July \or August \or September \or October \or November \or December \fi
  \number\year\timecount=\number\time
  \divide\timecount by 60}
\newdimen\strutdepth
\def\DRAFT{\def\lmargin(##1){\strut\vadjust{\kern-\strutdepth
  \vtop to \strutdepth{
  \baselineskip\strutdepth\vss\rlap{\kern-1.2 truecm\eightpoint{##1}}}}}
  \font\footfont=cmti7
  \footline={{\footfont \hfil File:\jobname, \TODAY,  \number\timecount h}}}
\def\lmargin(#1){}
\def\ifundefined#1{\expandafter\ifx\csname#1\endcsname\relax}
\def\ifff(#1,#2,#3){\ifundefined{#1#2}
  \expandafter\xdef\csname #1#2\endcsname{#3}\else
  \write16{Warning : doubly defined #1,#2}\fi}
\def\NEWDEF #1,#2,#3 {\ifff({#1},{#2},{#3})}
\def\EQ(#1){\lmargin(#1)\eqno\tag(#1)}
\def\NR(#1){&\lmargin(#1)\tag(#1)\cr}  
\def\tag(#1){({\rm \actualnumber.\number\EQNcount})
  \NEWDEF e,#1,(\actualnumber.\number\EQNcount)
  \global\advance\EQNcount by 1
  }
\def\equ(#1){\ifundefined{e#1}$\spadesuit$#1
  \else\csname e#1\endcsname\fi}
\def\CLAIM #1(#2) #3\par{
  \vskip.1in\medbreak\noindent
  {\lmargin(#2)\bf #1~\actualnumber.\number\CLAIMcount.} {\sl #3}\par
  \NEWDEF c,#2,{#1~\actualnumber.\number\CLAIMcount}
  \global\advance\CLAIMcount by 1
  \ifdim\lastskip<\medskipamount
  \removelastskip\penalty55\medskip\fi}
\def\CLAIMNONR #1 #2\par{
  \vskip.1in\medbreak\noindent
  {\bf #1.} {\sl #2}\par
  \ifdim\lastskip<\medskipamount
  \removelastskip\penalty55\medskip\fi}
\def\clm(#1){\ifundefined{c#1}$\spadesuit$#1\else\csname c#1\endcsname\fi}
\def\sectionsize{\twelvepoint}
\def\sectiontype{\bf}
\newskip\beforesectionskipamount  
\newskip\sectionskipamount        
\beforesectionskipamount=24pt plus8pt minus8pt
\sectionskipamount=3pt plus1pt minus1pt
\newskip\beforesubsectionskipamount  
\newskip\subsectionskipamount        
\beforesubsectionskipamount=12pt plus4pt minus4pt
\subsectionskipamount=2pt plus1pt minus1pt
\def\sectionskip{\vskip\sectionskipamount}
\def\beforesectionskip{\vskip\beforesectionskipamount}
\def\subsectionskip{\vskip\subsectionskipamount}
\def\beforesubsectionskip{\vskip\beforesubsectionskipamount}
\def\SECTION#1\par{\vskip0pt plus.3\vsize\penalty-75
  \vskip0pt plus -.3\vsize
  \global\advance\SECTIONcount by 1
  \def\actualnumber{\number\SECTIONcount}
  \beforesectionskip\noindent
  {\sectionsize\sectiontype \actualnumber.\ #1}
  \EQNcount=1
  \CLAIMcount=1
  \SUBSECTIONcount=0
  \nobreak\sectionskip\noindent}
\def\SECTIONNONR#1\par{\vskip0pt plus.3\vsize\penalty-75
  \vskip0pt plus -.3\vsize
  \global\advance\SECTIONcount by 1
  \beforesectionskip\noindent
  {\sectionsize\sectiontype #1}
  \EQNcount=1
  \CLAIMcount=1
  \SUBSECTIONcount=0
  \nobreak\sectionskip\noindent}
\def\SECT(#1)#2\par{\lmargin(#1)
  \SECTION #2\par
  \NEWDEF s,#1,{\actualnumber}}
\def\sec(#1){\ifundefined{s#1}$\spadesuit$#1
  \else Section \csname s#1\endcsname\fi}
\def\subsectionsize{}
\def\subsectiontype{\bf}
\def\SUBSECTION#1\par{\vskip0pt plus.2\vsize\penalty-75
  \vskip0pt plus -.2\vsize
  \global\advance\SUBSECTIONcount by 1
  \beforesubsectionskip\noindent
  {\subsectionsize\subsectiontype \actualnumber.\number\SUBSECTIONcount.\ #1}
  \nobreak\subsectionskip\noindent}
\def\SUBSECT(#1)#2\par{\lmargin(#1)
  \SUBSECTION #2\par
  \NEWDEF p,#1,{\actualnumber.\number\SUBSECTIONcount}}
\def\subsec(#1){\ifundefined{p#1}$\spadesuit$#1
  \else Subsection \csname p#1\endcsname\fi}
\def\pap{\rm}
\def\bok{\sl}
\newcount\BIBflag \BIBflag=0
\newcount\BIBcount \BIBcount=0
\def\Item#1#2\par{%
  \global\advance\BIBcount by 1
  \ifnum\BIBflag=0 \NEWDEF b,#1,[\number\BIBcount] 
  \else \item{[\number\BIBcount]}\lmargin({#1})#2\par\fi
  \ifundefined{q#1}
    \ifnum\BIBflag=1\write16{Warning : unquoted reference #1}\fi\fi}
\def\ref#1{%
  \ifundefined{b#1}$\spadesuit$#1\else\csname b#1\endcsname\fi
  \ifundefined{q#1}\expandafter\xdef\csname q#1\endcsname{}\fi}
\def\REFERENCES{\BIBflag=1 \BIBcount=0
  \parindent=20pt
  \parskip=2pt  
  \SECTIONNONR References\par}

\def\APPENDIX(#1)#2\par{
  \def\actualnumber{#1}
  \SECTIONNONR Appendix #1. #2\par}
\def\PROOF{\medskip\noindent{\bf Proof.\ }}
\def\REMARK{\medskip\noindent{\bf Remark.\ }}
\def\REMARKS{\medskip\noindent{\bf Remarks.\ }}

\def\LIKEREMARK(#1){\medskip\noindent{\bf #1.\ }}
%
%
\let\endarg=\par
\def\finish{\def\endarg{\par\endgroup}}
\def\start{\endarg\begingroup}

  \def\beginFROM{\start\parskip=0pt\vskip\baselineskip
    \def\finish{\def\endarg{\egroup\par\endgroup}}
    \vbox\bgroup\obeylines\eightpoint\em\finish}
  
\def\ABSTRACT#1\par{
  \vskip 1in {\noindent\sectionsize\sectiontype Abstract.} #1 \par}

%
%
\newdimen\texpscorrection
\texpscorrection=0truecm  
\newcount\FIGUREcount \FIGUREcount=0
\newskip\ttglue 
\newdimen\figcenter
\def\figure #1 #2 #3 #4\cr{\null
  \global\advance\FIGUREcount by 1
  \NEWDEF fig,#1,{Fig.~\number\FIGUREcount}
  \write16{ FIG \number\FIGUREcount: #1}
  {\goodbreak\figcenter=\hsize\relax
  \advance\figcenter by -#3truecm
  \divide\figcenter by 2
  \midinsert\vskip #2truecm\noindent\hskip\figcenter
  \includegraphics{#1}
  \vskip 0.8truecm\noindent \vbox{\eightpoint\noindent
  {\bf\fig(#1)}: #4}\endinsert}}
\def\figurewithtex #1 #2 #3 #4 #5\cr{\null
  \global\advance\FIGUREcount by 1
  \NEWDEF fig,#1,{Fig.~\number\FIGUREcount}
  \write16{ FIG \number\FIGUREcount: #1}
  {\goodbreak\figcenter=\hsize\relax
  \advance\figcenter by -#4truecm
  \divide\figcenter by 2
  \midinsert\vskip #3truecm\noindent\hskip\figcenter
  \includegraphics{#1}{\hskip\texpscorrection\input #2 }
  \vskip 0.8truecm\noindent \vbox{\eightpoint\noindent
  {\bf\fig(#1)}: #5}\endinsert}}
\def\fig(#1){\ifundefined{fig#1}$\spadesuit$#1
  \else\csname fig#1\endcsname\fi}
%
%
\catcode`@=11
\def\footnote#1{\let\@sf\empty 
  \ifhmode\edef\@sf{\spacefactor\the\spacefactor}\/\fi
  #1\@sf\vfootnote{#1}}
\def\vfootnote#1{\insert\footins\bgroup\eightpoint
  \interlinepenalty\interfootnotelinepenalty
  \splittopskip\ht\strutbox 
  \splitmaxdepth\dp\strutbox \floatingpenalty\@MM
  \leftskip\z@skip \rightskip\z@skip \spaceskip\z@skip \xspaceskip\z@skip
  \textindent{#1}\footstrut\futurelet\next\fo@t}
\def\fo@t{\ifcat\bgroup\noexpand\next \let\next\f@@t
  \else\let\next\f@t\fi \next}
\def\f@@t{\bgroup\aftergroup\@foot\let\next}
\def\f@t#1{#1\@foot}
\def\@foot{\strut\egroup}
\def\footstrut{\vbox to\splittopskip{}}
\skip\footins=\bigskipamount 
\count\footins=1000 
\dimen\footins=8in  
\catcode`@=12       
%
\font\twelverm=cmr12
\font\twelvei=cmmi12
\font\twelvesy=cmsy10 scaled\magstep1
\font\twelveex=cmex10 scaled\magstep1
\font\twelvebf=cmbx12 
\font\twelvett=cmtt12
\font\twelvesl=cmsl12
\font\twelveit=cmti12
\font\ninerm=cmr9

\font\ninesy=cmsy9

\font\eightrm=cmr8
\font\eighti=cmmi8
\font\eightsy=cmsy8
\font\eightex=cmex8
\font\eightbf=cmbx8
\font\eighttt=cmtt8
\font\eightsl=cmsl8
\font\eightit=cmti8
\font\sixrm=cmr6
\font\sixi=cmmi6
\font\sixsy=cmsy6
\font\sixbf=cmbx6
\newfam\truecmr 
\newfam\truecmsy
\font\twelvetruecmr=cmr10 scaled\magstep1
\font\twelvetruecmsy=cmsy10 scaled\magstep1
\font\tentruecmr=cmr10
\font\tentruecmsy=cmsy10
\font\eighttruecmr=cmr8
\font\eighttruecmsy=cmsy8
\font\seventruecmr=cmr7
\font\seventruecmsy=cmsy7
\font\sixtruecmr=cmr6
\font\sixtruecmsy=cmsy6
\font\fivetruecmr=cmr5
\font\fivetruecmsy=cmsy5
\textfont\truecmr=\tentruecmr
\scriptfont\truecmr=\seventruecmr
\scriptscriptfont\truecmr=\fivetruecmr
\textfont\truecmsy=\tentruecmsy
\scriptfont\truecmsy=\seventruecmsy
\scriptscriptfont\truecmsy=\fivetruecmsy
%
\def \eightpoint{\def\rm{\fam0\eightrm}
  \textfont0=\eightrm \scriptfont0=\sixrm \scriptscriptfont0=\fiverm 
  \textfont1=\eighti  \scriptfont1=\sixi  \scriptscriptfont1=\fivei 
  \textfont2=\eightsy \scriptfont2=\sixsy \scriptscriptfont2=\fivesy 
  \textfont3=\eightex \scriptfont3=\eightex \scriptscriptfont3=\eightex
  \textfont\itfam=\eightit          \def\it{\fam\itfam\eightit}%
  \textfont\slfam=\eightsl          \def\sl{\fam\slfam\eightsl}%
  \textfont\ttfam=\eighttt          \def\tt{\fam\ttfam\eighttt}%
  \textfont\bffam=\eightbf          \scriptfont\bffam=\sixbf
  \scriptscriptfont\bffam=\fivebf   \def\bf{\fam\bffam\eightbf}%
  \textfont\truecmr=\eighttruecmr   \scriptfont\truecmr=\sixtruecmr
  \scriptscriptfont\truecmr=\fivetruecmr
  \textfont\truecmsy=\eighttruecmsy \scriptfont\truecmsy=\sixtruecmsy
  \scriptscriptfont\truecmsy=\fivetruecmsy
  \tt \ttglue=.5em plus.25em minus.15em 
  \setbox\strutbox=\hbox{\vrule height7pt depth2pt width0pt}%
  \normalbaselineskip=9pt
  \let\sc=\sixrm  \let\big=\eightbig  \normalbaselines\rm
}
\def \twelvepoint{\def\rm{\fam0\twelverm}
\textfont0=\twelverm  \scriptfont0=\tenrm  \scriptscriptfont0=\eightrm 
\textfont1=\twelvei   \scriptfont1=\teni   \scriptscriptfont1=\eighti 
\textfont2=\twelvesy  \scriptfont2=\tensy  \scriptscriptfont2=\eightsy 
\textfont3=\twelveex  \scriptfont3=\tenex  \scriptscriptfont3=\eightex 
\textfont\itfam=\twelveit             \def\it{\fam\itfam\twelveit}%
\textfont\slfam=\twelvesl             \def\sl{\fam\slfam\twelvesl}%
\textfont\ttfam=\twelvett             \def\tt{\fam\ttfam\twelvett}%
\textfont\bffam=\twelvebf             \scriptfont\bffam=\tenbf
\scriptscriptfont\bffam=\eightbf      \def\bf{\fam\bffam\twelvebf}%
\textfont\truecmr=\twelvetruecmr      \scriptfont\truecmr=\tentruecmr
\scriptscriptfont\truecmr=\eighttruecmr
\textfont\truecmsy=\twelvetruecmsy    \scriptfont\truecmsy=\tentruecmsy
\scriptscriptfont\truecmsy=\eighttruecmsy
\tt \ttglue=.5em plus.25em minus.15em 
\setbox\strutbox=\hbox{\vrule htwelve7pt depth2pt width0pt}%
\normalbaselineskip=12pt
\let\sc=\tenrm  \let\big=\twelvebig  \normalbaselines\rm
}
\catcode`@=11
\def\eightbig#1{{\hbox{$\textfont0=\ninerm\textfont2=\ninesy\left#1
  \vbox to6.5pt{}\right.\n@space$}}}
\catcode`@=12
%
%

\def\CC{{\cal C}}

\def\EE{{\cal E}}

\def\LL{{\cal L}}

\def\NN{{\cal N}}
\def\OO{{\cal O}}

\def\XX{{\cal X}}
\def\YY{{\cal Y}}
\def\ZZ{{\cal Z}}

\def\HB {\hfill\break}
\def\sqr#1#2{{\vcenter{\vbox{\hrule height .#2pt
        \hbox{\vrule width.#2pt height#1pt \kern#1pt
           \vrule width.#2pt}
        \hrule height.#2pt}}}}
\def\square{\mathchoice\sqr64\sqr64\sqr{2.1}3\sqr{1.5}3}        
\def\QED{\null\hfill$\square$}

\def\eqdef{\buildrel\hbox{\sevenrm def}\over =}
\def\real{{\bf R}}
\def\natural{{\bf N}}
\def\complex{{\bf C}}

\def\Re{{\rm Re\,}}

\def\epsilon{\varepsilon}
\def\phi{\varphi}
\def\lbar{\underline{\ell}}
\def\pp{{\sl a.e.}}
\def\L{{\rm L}}
\def\H{{\rm H}}
\def\loc{{\rm loc}}
\def\e{{\rm e}}

\def\d{{\rm \,d}}
\def\dd{{\rm d}}
\def\frac#1#2{{\textstyle{#1 \over #2}}}
%
%
\font\title=cmbx10 at 14pt
\centerline{\title Stability of Propagating Fronts}
\vskip 2mm
\centerline{\title in Damped Hyperbolic Equations}

\baselineskip=12pt
\vskip 1.5cm
\noindent{\bf Th. Gallay, G. Raugel}\HB 
Analyse Num\'erique et EDP\HB 
CNRS et Universit\'e de Paris-Sud\HB
F-91405 Orsay Cedex, France\HB
{\tt Thierry.Gallay@math.u-psud.fr}\HB
{\tt Genevieve.Raugel@math.u-psud.fr}

\baselineskip=\normalbaselineskip
\vskip 5mm
\noindent{\bf Abstract}. -- {\sl We consider the damped hyperbolic
equation
$$
   \epsilon u_{tt} + u_t \,=\, u_{xx} + F(u)~, \quad x \in \real~, 
   \quad t \ge 0~, \quad u \in \real~,
$$
where $\epsilon$ is a positive, not necessarily small parameter. We
assume that $F(0)=F(1)=0$ and that $F$ is concave on the interval $[0,1]$. 
Under these assumptions, our equation has a continuous family of monotone
propagating fronts (or travelling waves) indexed by the speed parameter 
$c \ge c_*$. Using energy estimates, we first show that the travelling waves 
are locally stable with respect to perturbations in a weighted Sobolev space. 
Then, under additional assumptions on the non-linearity, we obtain global 
stability results using a suitable version of the hyperbolic Maximum 
Principle. Finally, in the critical case $c = c_*$, we use self-similar
variables to compute the exact asymptotic behavior of the perturbations
as $t \to +\infty$. In particular, setting $\epsilon = 0$, we recover
several stability results for the travelling waves of the corresponding
parabolic equation.}


\vskip 1mm
\noindent AMS classification codes (1991) : 35B40, 35B35, 35B30, 35L05, 35C20

\vfill\eject


\SECTION Introduction

Mathematical models for spreading and interacting particles or individuals
are very common in chemistry and biology, especially in genetics and 
population dynamics. If the spatial spread of the particles is described
by Brownian motion, these models usually take the form of reaction-diffusion
equations or systems for the population densities [10,27].
Depending on the precise form of the interaction, such systems exhibit 
interesting solutions like propagating fronts or travelling waves, whose 
existence and stability properties have attracted a lot of attention in 
recent years [32]. 

It can be argued, however, that diffusion is not a realistic model
of spatial spread for short times, because particles performing 
Brownian motion can move with arbitrarily high speed, and the directions
of motion at successive times are uncorrelated. This drawback can
be eliminated by replacing the diffusion process with a velocity jump
process which is more satisfactory for short times and has equivalent
long-time properties [17,22,31]. 
In one space dimension, this procedure leads to damped wave equations 
instead of reaction-diffusion systems [6,19,20,33]. Under the same 
assumptions as in the parabolic case, the damped hyperbolic equations 
also have travelling wave solutions [18] with analogous stability properties 
[14,16]. The aim of this paper is to review some of these stability results 
in the simplest case of a scalar equation with a non-linearity of 
``monostable'' type. 

We thus consider the damped hyperbolic equation
$$
   \epsilon u_{tt} + u_t \,=\, u_{yy} + F(u)~, \EQ(u)
$$
where $y \in \real$, $t \in \real_+$ and $\epsilon$ is a positive, {\sl not
necessarily small} parameter. We assume that the non-linearity $F \in 
\CC^2(\real,\real)$ has the following properties:
$$
   F(0) = F(1) = 0~, \quad F'(0) > 0~, \quad F'(1) < 0~, \quad
   F''(u) < 0 ~~\hbox{for } u \in (0,1]~. 
   \leqno({\rm H1})
$$
In particular, $u \equiv 1$ is a stable equilibrium point of Eq.\equ(u), 
and $u \equiv 0$ is unstable. For simplicity, we assume $F$ being concave 
on $[0,1]$, but this condition is more restrictive than what we really
need and could be relaxed in several ways. A typical non-linearity 
satisfying (H1) is $F(u) = u-u^m$, with $m \in \natural$, $m \ge 2$. 

Under the assumptions (H1), Eq.\equ(u) has monotone travelling wave solutions
(also called propagating fronts) connecting the equilibrium states $u=1$ and 
$u=0$. Indeed, setting
$$
   u(y,t) \,=\, h(\sqrt{1+\epsilon c^2}y - ct) \,\equiv\, h(x)~, \EQ(TW)
$$
where $c > 0$, we obtain for the function $h$ the ordinary differential 
equation
$$
   h''(x) + c h'(x) + F(h(x)) \,=\, 0~, \quad x \in \real~. \EQ(front)
$$
As is well-known [1,26], Eq.\equ(front) has a solution 
satisfying $h'(x) < 0$ for all $x \in \real$, $h(-\infty) = 1$, 
$h(+\infty) = 0$
if and only if $c \ge c_* = 2\sqrt{F'(0)}$. This solution is unique 
up to a translation in the variable $x$. Therefore, for all $\epsilon > 0$,
Eq.\equ(u) has a continuous family of monotone travelling waves indexed
by the speed parameter $c \ge c_*$. Note that the actual
speed of the wave is not $c$, but $c/\sqrt{1{+}\epsilon c^2}$, a quantity
which is bounded by $1/\sqrt{\epsilon}$ for all $c \ge c_*$. 

In the limit $\epsilon \to 0$, Eq.\equ(u) reduces to the semilinear 
parabolic equation $u_t = u_{yy} + F(u)$, which has been intensively 
studied since the pioneering work of Fisher [11] and Kolmogorov,
Petrovskii and Piskunov [26]. Using the parabolic Maximum Principle and 
probabilistic techniques, the long-time behavior of a large class
of solutions has been explicitly determined [1,2]. 
In a more general context, a local stability analysis of the travelling
waves using functional-analytic methods has been initiated by Sattinger
[30] and extended by many authors [7,23,25].
In particular, the decay rate in time of the perturbations in the 
critical case $c = c_*$ has been computed using renormalization 
techniques [3,13]. Since the equilibrium state $u=0$ ahead
of the front is linearly unstable, all these stability properties are
restricted to perturbations which decay to zero at least as fast as 
$h$ itself as $x \to +\infty$. Following [14,16], the aim of this paper 
is to show how these results can be extended to the hyperbolic case 
$\epsilon > 0$. 

To investigate the stability of the travelling wave \equ(TW) as a solution
of \equ(u), we go to a moving frame using the change of variables
$$
   u(y,t) \,=\, v(\sqrt{1+\epsilon c^2}y - ct,t) \,\equiv
   v(x,t)~,
$$
where $x = \sqrt{1+\epsilon c^2}y - ct$. The equation for $v$ is
$$
   \epsilon v_{tt} + v_t - 2\epsilon c v_{xt} \,=\, 
   v_{xx} + c v_x + F(v)~, \EQ(v)
$$
and, by construction, $h(x)$ is a stationary solution of \equ(v).
Setting  $v(x,t) = h(x) + w(x,t)$, we obtain for the perturbation $w$
the equation
$$
   \epsilon w_{tt} + w_t - 2\epsilon c w_{xt} \,=\, 
   w_{xx} + c w_x + F'(h)w + \NN(h,w)w^2~, \EQ(w) 
$$
where
$$
   \NN(h,w) \,=\, \int_0^1 (1-\sigma) F''(h+\sigma w) \d\sigma~.
$$
Rewriting in the usual way the second order equation \equ(w) as a first order 
system for the pair $(w,w_t)$, we shall study the stability of the origin 
$(w,w_t) = (0,0)$ in an weighted Sobolev space $Z$ which we now describe. 

For $k \in \natural$, we denote by $\H^k = \H^k(\real)$ the usual (real)
Sobolev space of order $k$ over $\real$, with $\H^0(\real) = \L^2(\real)$.
Following [30], we introduce for $s > 0$ the weight function 
$p_s(x) = 1+\e^{sx}$ and, for $k \in \natural$, we denote by $\H^k_s$ the 
Hilbert space $\H^k(\real,p_s^2 \d x)$ defined by the norm
$$
   \|u\|_{\H^k_s}^2 \,=\, \int_\real \left(\sum_{i=0}^k 
   |\partial_x^i u(x)|^2\right) p_s(x)^2 \d x~. \EQ(Hnorm)
$$
Setting $\L^2_s = \H^0_s$, we define the product space $Z = \H^1_s 
\times \L^2_s$ equipped with the norm
$$
   \|(w_1,w_2)\|_Z^2 \,=\, \|w_1\|_{\H^1_s}^2 + \|w_2\|_{\L^2_s}^2~.
   \EQ(Znorm) 
$$
Finally, it will be convenient to denote by $Z_\epsilon$ the space $Z$ 
equipped with the $\epsilon$-dependent norm
$$
   \|(w_1,w_2)\|_{Z_\epsilon}^2 \,=\, \|w_1\|_{\H^1_s}^2 + \epsilon 
   \|w_2\|_{\L^2_s}^2~. \EQ(Zepsnorm) 
$$

The perturbation space $Z$ clearly depends on the choice of $s > 0$ and 
becomes smaller when $s$ increases. It is thus natural to look for the 
the smallest value of $s > 0$ for which the origin in \equ(w) is linearly 
stable in $Z$. This is most conveniently done by setting $w(x,t) = \e^{-sx}
\omega(x,t)$ and studying the equation for $\omega$, namely:
$$\eqalign{
   \epsilon \omega_{tt} &+ (1+2\epsilon cs)\omega_t - 2\epsilon c 
   \omega_{xt} \,=\, \cr
   & \omega_{xx} + (c-2s) \omega_x + (F'(h)-cs+s^2)
   \omega + \NN(h,\e^{-sx}\omega)\e^{-sx} \omega^2~.} \EQ(omega)
$$
A straightforward computation in the Fourier variables shows that the
origin in \equ(omega) is linearly stable in $\H^1\times\L^2$ only if 
$F'(0) - cs + s^2 \le 0$. In fact, this condition can easily be inferred from 
the coefficient of $\omega$ in \equ(omega). Therefore, the largest 
perturbation space $Z$ for which we can expect stability of the front $h$ 
is obtained by choosing
$$
   s \,=\, {1 \over 2} (c - \sqrt{c^2 - c_*^2})~. \EQ(schoice)
$$
Note that this value corresponds to the decay 
rate of $h$ as $x \to +\infty$, since $h(x) \sim \e^{-sx}$ if $c > c_*$ 
and $h(x) \sim x \e^{-sx}$ if $c = c_*$ [1]. Thus, if \equ(schoice)
holds, the translations $h(x+x_0)-h(x)$ do not belong to the perturbation 
space $\H^1_s$. In view of the translation invariance of Eq.\equ(u), this is
of course necessary to obtain an asymptotic stability result. In the 
sequel, we always assume that the condition \equ(schoice) holds. 

In this paper, we present three different results which show that the
travelling wave $h$ is stable with respect to perturbations in $Z_\epsilon$ 
or in a subspace of it. 
In Section~2, we give a local stability result valid for all $c \ge 
c_*$ and all $\epsilon > 0$. Using appropriate energy functionals, we show 
that, if $(w(0),w_t(0))$ is sufficiently small in $Z_\epsilon$, then the 
solution $(w(t),w_t(t))$ of \equ(w) stays in a neighborhood of the origin in 
$Z_\epsilon$ and converges to zero as $t \to +\infty$ in a slightly weaker 
norm. In the critical case $c = c_*$, we also give an estimate of the 
convergence rate. These results have been proved in [14] in the 
particular case where $F(u) = u-u^2$. Their proofs can be adapted, with 
minor changes, to cover the general case of a non-linearity $F$ satisfying 
(H1).

Section~3 is devoted to global stability results. We first recall
the Maximum Principle for hyperbolic equations [29] in a version 
adapted to our problem. Then, under the additional assumption that $F'(u)$ 
be strictly negative for $u \ge 1$, we show that the travelling wave $h$ is 
stable with respect to ``large'' perturbations in $Z_\epsilon$, provided some
positivity conditions are satisfied. Furthermore, if $1+4\epsilon F'(1) 
\ge 0$ and $F''(u) \le 0$ for $u \ge 0$, we obtain linear upper and lower 
bounds for the solutions of Eq.\equ(w), as well as a decay rate in time for 
the quantity $\|p_s w(t)\|_{\L^\infty}$. The proofs of these results can also 
be found in [14] if $F(u) = u-u^2$. 

In Section~4, we restrict our analysis to the critical case $c=c_*$, and
we study the long-time behavior of the solutions $(w,w_t)$ of \equ(w)
in a slightly smaller function space. In particular, we show that
$$
   w(x,t) \,=\, {\alpha \over t^{3/2}} \,h'(x)\phi^*\left({x
   \sqrt{1{+}\epsilon c_*^2} \over \sqrt{t}}\right) + o(t^{-3/2})~, 
   \quad t \to +\infty~,
$$
where $\alpha \in \real$ and $\phi^* : \real \to \real$ is a {\sl 
universal} profile. In the parabolic case $\epsilon = 0$, this asymptotic
expansion has been obtained by Gallay [13] using the renormalization
group method [5] combined with resolvent estimates. 
We follow here a simpler and fairly different approach based on
self-similar variables and energy estimates only. We refer to [16]
for the detailed proof. 

To conclude this introduction, we would like to point out the striking 
similarity between the stability results presented here and the corresponding
statements in the parabolic case. This is an illustration of the more
general fact that the long-time behavior of solutions to damped hyperbolic 
equations such as \equ(u) is essentially parabolic [15]. 
A similar phenomenon can be observed in the context of hyperbolic 
conservation laws with damping [21,28].


\SECTION Local Stability of the Travelling Waves

In this section, we show that the travelling wave $h$ is stable with respect 
to sufficiently small perturbations in the space $Z_\epsilon$. Our main 
result is:

\CLAIM Theorem(local) Assume that (H1) holds, and let $\epsilon_0 > 0$, 
$c \ge c_*$. Then there exist 
constants $\delta_0 > 0$ and $K_0 \ge 1$ such that, for all $0 < \epsilon 
\le \epsilon_0$, the following result holds: for all $(\phi_0,\phi_1) \in 
Z_\epsilon$ such that $\|(\phi_0,\phi_1)\|_{Z_\epsilon} \le \delta_0$, there 
exists a unique solution $(w,w_t) \in \CC^0([0,\infty),Z_\epsilon)$ of
\equ(w) with initial data $(w(0),w_t(0))=(\phi_0,\phi_1)$. Moreover, one has
$$
   \|(w(t),w_t(t))\|_{Z_\epsilon} \,\le\, K_0 
   \|(\phi_0,\phi_1)\|_{Z_\epsilon}~, \quad t \ge 0~, \EQ(globound)
$$
and
$$
   \lim_{t \to +\infty} \left(\|w(t)\|_{\H^1} + \|(p_s w(t))_x\|_{\L^2}
   + \|p_s w_t(t)\|_{\L^2}\right) \,=\, 0~, \EQ(tozero)
$$
where $p_s(x) = 1+\e^{sx}$ and s is given by \equ(schoice). 

\noindent{\bf Sketch of the proof.} We follow the lines of the proof of
Theorem~1.1 in [14]. Let
$$
  \NN_1(h) \,=\, \int_0^1 F''(\sigma h)\d\sigma~, \quad
  \NN_2(h,w) \,=\, {1 \over 2} \int_0^1 (1{-}\sigma)^2 F''(h+\sigma w)
  \d\sigma~.
$$
Given $0 < \epsilon \le \epsilon_0$ and $c \ge c_*$, we assume that 
$(w,w_t) \in \CC^0([0,T],Z_\epsilon)$ is a solution of \equ(w) satisfying
$$
   \|w(t)\|_{\H^1_s} \,\le\, \delta~, \quad t \in [0,T]~, \leqno({\rm A1})
$$
for some (sufficiently small) $\delta > 0$. As in \equ(omega), we set $w(x,t) 
= \e^{-sx} \omega(x,t)$. To control the behavior of $w$ on $[0,T]$, we 
introduce two families of energy functionals:
$$ \eqalign{
   E_0(t) \,&=\, {1 \over 2} \int_\real \left(\epsilon \omega_t^2 + 
     \omega_x^2 -\omega^2 (h\NN_1(h) +2w\NN_2(h,w)) \right) \d x~, \cr
   E_1(t) \,&=\, \int_\real \left( (\frac12+\epsilon cs)\omega^2 + 
     \epsilon \omega \omega_t \right)\d x~, \cr
   E_2(t) \,&=\, \alpha_0 E_0(t) + E_1(t)~,}
$$ 
where $\alpha_0 = \max(2\epsilon,1/(2c^2))$, and
$$ \eqalign{
   \EE_0(t) \,&=\, {1 \over 2}\int_\real \left(\epsilon w_t^2 + 
     w_x^2 + s^2 w^2 -w^2 (h\NN_1(h)+2w\NN_2(h,w)) \right) \d x~, \cr
   \EE_1(t) \,&=\, \int_\real \left( (\frac12-\epsilon cs) w^2 + 
     \epsilon w w_t \right) \d x~, \cr
   \EE_2(t) \,&=\, \alpha_1 \EE_0(t) + \EE_1(t) + \alpha_2 (
     E_0(t)E_2(t))^{1/2}~,}
$$
where $\alpha_1 = \max(2\epsilon,\theta/(2c^2))$ and $\theta,\alpha_2$ are 
positive constants. Note that the functionals $E_i$, $\EE_i$ control
the behavior of the perturbation $w$ ``ahead'' and ``behind'' the front 
respectively. 

Arguing as in [14, Section~2], one can show that, if $\delta$, $\theta$ and 
$\alpha_2^{-1}$ are sufficiently small, then the functions $E_0$, $E_2$, 
$\EE_2$ are non-negative and satisfy the differential inequalities
$$\eqalign{
   &{\dd E_0 \over \dd t}(t) \,\le\, {1 \over 2}(c^2-c_*^2)\,E_0(t)~, \quad
    {\dd E_2 \over \dd t}(t) + E_0(t) \,\le\, 0~, \cr
   &{\dd \EE_2 \over \dd t}(t) + \alpha_3 \EE_2(t) \,\le\, C_1 
   (E_0(t)E_2(t))^{1/2}~, \quad t \in [0,T]~,} \EQ(diff)
$$
for some $\alpha_3, C_1 > 0$. In addition, there exists a constant $C_2 \ge 1$
such that
$$
  C_2^{-1} \|(w,w_t)\|_{Z_\epsilon}^2 \,\le\, E_2(t) + \EE_2(t) 
  \,\le\, C_2 \|(w,w_t)\|_{Z_\epsilon}^2 (1 + \Psi(\|w(t)\|_{\L^\infty}))~,
  \EQ(bounds)
$$
where the function $\Psi : \real_+ \to \real_+$ is defined by
$$
   \Psi(K) \,=\, \sup_{0 \le u \le 1{+}K} |F'(u)|~. \EQ(PSI)
$$
Combining the estimates Eqs.\equ(diff), \equ(bounds), we obtain the bound
$$
   \|(w(t),w_t(t))\|_{Z_\epsilon} \,\le\, C_0 
   \|(w(0),w_t(0))\|_{Z_\epsilon} (1 + \Psi(\|w(0)\|_{\L^\infty}))^{1/2}~,
   \quad t \in [0,T]~, \EQ(gbound)
$$
where $C_0$ is a positive constant depending only on $\epsilon_0$, $c$ and
$F$. Since the Cauchy problem for Eq.\equ(w) in $Z_\epsilon$ is locally 
well-posed [14], this proves global existence of the solution $(w,w_t)$ 
provided the right-hand side of \equ(gbound) is smaller than the quantity 
$\delta$ appearing in (A1). Then, the differential inequalities \equ(diff) 
imply that $E_0(t)$ and $\EE_2(t)$ converge to zero as $t \to +\infty$, thus 
proving \equ(tozero). \QED

\REMARKS\HB
{\bf 1)} In the critical case $c=c_*$, it follows from \equ(diff) 
that $E_0$ is non-increasing in time, and that $tE_0(t) + 
t^{1/2}\EE_2(t) \to 0$ as $t \to +\infty$. In particular, we have
$$
   \lim_{t \to +\infty} \left(t^{1/4}(\|w(t)\|_{\H^1} + \|w_t(t)\|_{\L^2})
   + t^{1/2}(\|\omega_x(t)\|_{\L^2} + \|\omega_t(t)\|_{\L^2})\right)
   \,=\, 0~. \EQ(rate)
$$
This decay rate is not optimal, see Section~4 below. 

\noindent{\bf 2)} Combining \equ(globound), \equ(tozero), \equ(rate), 
we obtain
$$
   \lim_{t \to +\infty} \|p_s w(t)\|_{\L^\infty} \,=\, 0~~\hbox{if }
   c > c_*~, \quad \hbox{and}\quad \lim_{t \to +\infty} t^{1/4}
   \|p_s w(t)\|_{\L^\infty} \,=\, 0~~\hbox{if } c = c_*~. \EQ(rates)
$$

\noindent{\bf 3)} All the estimates in the proof of \clm(local) are
uniform in $\epsilon$ for $\epsilon \in (0,\epsilon_0]$. In particular, 
taking the limit $\epsilon \to 0$ everywhere, we obtain a proof of
the local stability of the travelling wave $h$ for the corresponding 
parabolic equation. 

\noindent{\bf 4)} If $c=c_*$, the stability condition $F'(0)-cs+s^2 = 
(s-c_*/2)^2 \le 0$ implies $s = c_*/2$, hence \equ(schoice) is the
only possibility. On the other hand, if $c > c_*$ and $s$ is chosen so that 
$F'(0) -cs + s^2 < 0$ (in contrast to \equ(schoice)), one can show
using the same spectral estimates as in the parabolic case that
the origin in \equ(w) is exponentially stable in $Z_\epsilon$. The fastest 
decay rate in time for the perturbations is obtained if we set 
$s = \hat s(\epsilon)$, where
$$
   \hat s(\epsilon) \,=\, {c \over 2} \sqrt{{1+4\epsilon F'(0) \over 
   1+\epsilon c^2}}~.
$$


\SECTION Global Stability of the Travelling Waves

Throughout this section, we assume that the non-linearity $F$ satisfies
(H1) and
$$
   F'(u) \,\le\, -\mu < 0~, \quad u \ge 1~, \leqno({\rm H2})
$$
for some $\mu > 0$. Under this additional assumption, it is known 
in the parabolic case that the front $h$ is stable with 
respect to large perturbations in $\H^1_s$ satisfying a positivity
condition. This property is a consequence of the classical Maximum
Principle for parabolic equations. In this section, we show a similar
global stability result for $\epsilon > 0$ using a hyperbolic
Maximum Principle.  

Our starting point is the following crucial observation. Let $0 < \epsilon
\le \epsilon_0$, $c \ge c_*$, $d \in (0,1]$, and assume that $(w,w_t) \in 
\CC^0([0,T],Z_\epsilon)$ is a solution of \equ(w) satisfying, 
instead of (A1), 
$$
   w(x,t) \,\ge\, -(1-d)h(x)~, \quad x \in \real~, \quad t \in [0,T]~.
   \leqno({\rm A2})
$$
In other words, we assume that $v(x,t) \equiv h(x)+w(x,t) \ge dh(x)$. 
Then it can be shown (see [14] in the case $F(u) = u-u^2$) 
that the a priori estimates \equ(diff), \equ(bounds), \equ(gbound)
still hold for the solution $(w,w_t)$, with constants $C_0, C_1, C_2$ 
depending only on $\epsilon_0$, $c$, $d$ and $F$. Thus, we can remove 
the smallness condition (A1) in the proof of \clm(local) provided we
are able to show that the positivity condition (A2) is satisfied
for all times. This in turn can be obtained from the Maximum 
Principle under appropriate assumptions on the initial data.

\SUBSECTION The Hyperbolic Maximum Principle

Motivated by \equ(v) or \equ(w), we consider the hyperbolic operator $L$ with 
constant coefficients
$$
  L u \,=\, u_{xx} + 2 \epsilon c u_{xt} -\epsilon u_{tt} + c u_x - u_t~.
  \EQ(operator)
$$ 
Assume that $\ell: \real\times [0,T] \to \real$ is a continuous function 
such that
$$
   \ell(x,t) \,\ge\, \lbar~, \quad x \in \real~, \quad t \in [0,T]~,
  \EQ(ell)
$$
where $T$ is a positive number and $\lbar \in \real$ satisfies
$$
   1 + 4\epsilon\lbar \,\ge\, 0~. \EQ(lbar)
$$
The following result is a consequence of the Maximum Principle 
given by Protter and Weinberger (see [29, Chapter\ 4, Theorem\ 1]
or [14, Appendix\ A, Theorem\ A.1]). 

\CLAIM Theorem(MP) Let $\epsilon > 0$, $c > 0$, and assume that the 
conditions \equ(ell) and \equ(lbar) are satisfied. If $(u(x,t), u_t(x,t))$ 
belongs to $\CC^0([0,T], H_{\loc}^1(\real) \times L_{\loc}^2(\real))$, 
with $u_{xx} + 2\epsilon cu_{xt} -\epsilon u_{tt}$ in $L_{\loc}^2 
(\real \times (0,T))$, and if
$$
  \bigl(L+\ell(x,t)\bigr)u(x,t) \,\ge\, 0~, \quad \pp(x,t) \in \real 
  \times [0,T]~, \EQ(A4)
$$
$$
  u(x,0) \,\le\, 0~, \quad \forall \,x \in \real~, \EQ(A5)
$$
$$
  \epsilon u_t(x,0) - \epsilon c u_x (x,0) + {1\over 2}u(x,0) \,\le\, 0~,
  \quad \pp\hbox{ in }\real~, \EQ(A6)
$$
then $u(x,t) \le 0$ for all $(x,t) \in \real\times [0,T]$.

\noindent As an application, we define for $d \in [0,1]$, $K \ge 0$ the
function
$$
   \Lambda_d(K) \,=\, \inf\left\{ {F(v)-F(u) \over v-u} \,\Bigg|\,
   0 \le u \le d\,,~u < v \le 1{+}K \right\} \,\le\, 0~. \EQ(Lambda)
$$
We have the following result:

\CLAIM Proposition(41) Assume that (H1), (H2) hold. Let $\epsilon > 0$, 
$c \ge c_*$, $d \in [0,1]$ and let $K$ be a non-negative constant such that
$$
   1 + 4\epsilon \Lambda_d(K) \,\ge\, 0~. \EQ(41)
$$
For some $T > 0$, assume that $(w,w_t) \in \CC^0([0,T],Z_\epsilon)$
is a solution of \equ(w) with initial data $(\phi_0,\phi_1)$ satisfying
$$
   \phi_0(x) \,\ge\, - (1-d)h(x)~, \quad x \in \real~, \EQ(phi0)
$$
$$
  \epsilon \phi_1(x) \,\ge\, \epsilon c (\phi_0'(x) +(1{-}d)h'(x))
  - {1\over 2} (\phi_0(x)+(1{-}d)h(x))~, \quad \pp\hbox{ in }\real~. 
  \EQ(phi1)
$$
Suppose moreover that
$$
   w(x,t) \le K~, \quad (x,t) \in \real \times [0,T]~. \EQ(Kbound)
$$
Then
$$ 
  w(x,t) \ge  -(1-d)h(x)~,\quad (x,t) \in \real \times [0,T]~. \EQ(conclw)
$$

\PROOF Without loss of generality, we may assume that $F'(u) \ge 0$
for all $u \le 0$. Let $u(x,t) = dh(x)-v(x,t)$, where $v(x,t) = h(x) + 
w(x,t)$. Since $v(x,t)$ is a solution of \equ(v), it is straightforward 
to verify that $(L+\ell)u(x,t) = F(dh(x))-dF(h(x))$, where $L$ is defined 
in \equ(operator) and
$$
   \ell(x,t) \,=\, {F(v(x,t)) - F(dh(x)) \over v(x,t) -dh(x)}~.
$$
In view of (H1), one has $F(dh(x))-dF(h(x)) \ge 0$ for all $x \in \real$. 
Furthermore, since $v(x,t) \le 1+K$ by \equ(Kbound), we have $\ell(x,t) 
\ge \lbar = \Lambda_d(K)$. Indeed, this inequality follows immediately from 
\equ(Lambda) if $v(x,t) \ge dh(x)$; in the converse case, we observe
that $\ell(x,t) \ge \min(F'(dh(x)),0) \ge \Lambda_d(K)$. Thus \equ(41) 
implies \equ(lbar), and the conditions \equ(phi0), \equ(phi1) are nothing 
else as the hypotheses \equ(A5) and \equ(A6). Therefore \clm(MP) shows that 
$u(x,t) = dh(x)-v(x,t) \le 0$ for all $(x,t) \in \real \times [0,T]$, which 
is \equ(conclw). \QED

\REMARK \clm(MP) suggests the definition of a partial order in $H^1_{\loc}
(\real) \times L^2_{\loc}(\real)$ as follows. We say that $(\phi_0,\phi_1)
\le (\psi_0,\psi_1)$ if 
$$ \eqalign{
   \phi_0(x) \,&\le\, \psi_0(x)~, \quad x \in \real~, \cr
   \epsilon \phi_1(x) - \epsilon c \phi_0'(x) + {1 \over 2} \phi_0(x) \,&\le\, 
   \epsilon \psi_1(x) - \epsilon c \psi_0'(x) + {1 \over 2} \psi_0(x) 
   \quad \pp\hbox{ in }\real~,}
$$
see \equ(A5), \equ(A6). Then, if $(\phi_0,\phi_1) \le (\psi_0,\psi_1)$, 
the solution of the linear hyperbolic equation $(L+\ell)u(x,t) = 0$ satisfying 
$u(x,0) = \phi_0(x)$, $u_t(x,0) = \phi_1(x)$ stays for all $t \in \real_+$ 
below the solution of the same equation with initial data $(\psi_0,\psi_1)$. 
This order has the property that we can write any $(\phi_0,\phi_1) \in 
H^1_{\loc} \times L^2_{\loc}$ as the sum of a ``positive'' part 
$(\phi_0^+,\phi_1^+) \ge 0$ and a ``negative'' part $(\phi_0^-,\phi_1^-) 
\le 0$. This decomposition is unique if we impose that $(\phi_0^+,\phi_1^+) 
= 0$ whenever $(\phi_0,\phi_1) \le 0$ and $(\phi_0^-,\phi_1^-) = 0$ 
whenever $(\phi_0,\phi_1) \ge 0$. The formulas for $(\phi_0^\pm,\phi_1^\pm)$
are given by
$$ \eqalign{
   \phi_0^+(x) &= \sup(0,\phi_0(x))\,, \cr
   \phi_1^+(x) &= c (\phi_0^+)'(x) - {1 \over 2\epsilon} 
     \phi_0^+(x) + \sup(0,(\phi_1-c\phi_0' + {1 \over 2\epsilon} 
     \phi_0)(x))\,,} \EQ(phi+)
$$
and
$$ \eqalign{
   \phi_0^-(x) &= \inf(0,\phi_0(x))\,, \cr
   \phi_1^-(x) &= c (\phi_0^-)'(x) - {1 \over 2\epsilon} 
     \phi_0^-(x) + \inf(0,(\phi_1-c\phi_0' + {1 \over 2\epsilon}
     \phi_0)(x))\,.} \EQ(phi-)
$$
Remark that, if $(\phi_0,\phi_1) \in Z_\epsilon$, then $(\phi_0^\pm,
\phi_1^\pm) \in Z_\epsilon$. Moreover, it can be verified that
$|\phi_1^\pm(x)| \le |\phi_1(x)| + c |\phi_0'(x)|$ {\pp} in $\real$.  

\SUBSECTION A General Global Stability Result

Combining the a priori estimates of Section~2 and the Maximum Principle, 
we are now able to state and prove our main stability result:

\CLAIM Theorem(global) Assume that (H1), (H2) hold, and let 
$\epsilon_0 > 0$, $c \ge c_*$, $d \in (0,1]$. There exists a constant 
$C_0 \ge 1$ such that, for all $0 < \epsilon \le \epsilon_0$ and all 
$K > 0$ satisfying
$$
   1 + 4 \epsilon \Lambda_d(K) \,\ge\, 0~, \EQ(eps)
$$
the following result holds: If $K_* > 0$ is such that
$$
   C_0 K_* (1+\Psi(K_*))^{1/2} < K~, \EQ(Kstar)
$$
where $\Psi$ is defined in \equ(PSI), then for any $(\phi_0,\phi_1) \in 
Z_\epsilon$ satisfying the inequalities \equ(phi0), \equ(phi1) and the bound 
$\|(\phi_0,\phi_1)\|_{Z_\epsilon} \le K_*$, there exists a unique solution 
$(w,w_t) \in \CC^0([0,\infty),Z_\epsilon)$ of \equ(w) with initial data 
$(\phi_0,\phi_1)$. Moreover, one has 
$$ 
   \|(w(t),w_t(t))\|_{Z_\epsilon} \,\le\, K~, \quad
   w(x,t) \,\ge\, -(1-d)h(x)~, \EQ(lwbd)
$$
for all $x \in \real$, $t \in \real_+$, and \equ(tozero), \equ(rate) hold. 

\REMARK The constant $C_0$ in \equ(Kstar) is the same as in \equ(gbound). 

\noindent{\bf Sketch of the proof.} By \clm(41), the solution $(w,w_t)$
with initial data $(\phi_0,\phi_1)$ satisfies $w(x,t) \ge -(1-d)h(x)$ as 
long as $w(x,t) \le K$. On the other hand, in view of \equ(Kstar), the a 
priori estimate \equ(gbound) implies that $w(x,t) \le \|(w(t),w_t(t))
\|_{Z_\epsilon} < K$ as long as $w(x,t) \ge -(1-d)h(x)$. Combining
these facts and using a contradiction argument, we show that the solution
is globally defined and satisfies \equ(lwbd) for all times. The differential
inequalities \equ(diff) then imply \equ(tozero), \equ(rate). \QED

\REMARKS\HB
{\bf 1)} The relations \equ(eps), \equ(Kstar) imply that $K$ (hence $K_*$) 
can be chosen very big if $\epsilon$ is sufficiently small. In this case,   
\clm(global) shows that the travelling wave $h$ is stable with respect to 
large perturbations, provided the positivity conditions \equ(phi0), 
\equ(phi1) are satisfied. Conversely, if $\epsilon$ is large, then $K$ 
(hence $K_*$) has to be very small, and \clm(global) reduces to a local 
stability result similar to \clm(local).

\noindent{\bf 2)} As a simple example, consider the non-linearity 
$F(u) = u-u^2$ [14] which satisfies (H1), (H2). Then $\Lambda_d(K) = 
-(d+K)$, and the condition \equ(eps) reads $1-4\epsilon(d+K) \ge 0$. 
Also $\Psi(K) = 1+2K$, hence the hypothesis $\|(\phi_0,\phi_1)
\|_{Z_\epsilon} \le K_*$ can be replaced by
$$
   \|(\phi_0,\phi_1)\|_{Z_\epsilon} \le C K(1+K)^{-1/3}~,
$$
where $C$ is a sufficiently large positive constant. 

\SUBSECTION Further Stability Results if $1+4\epsilon F'(1) \ge 0$

The previous results are still incomplete for at least two reasons. 
First, if $c > c_*$, they fail to give a decay rate of the perturbations
as $t \to +\infty$, see \equ(rates). Next, they do not provide any global 
existence result if $d = 0$, that is if $v(x,0) \ge 0$. Assuming that the 
non-linearity $F$ satisfies the stronger assumption
$$
   F''(u) \,\le\, 0~, \quad u \,\ge\, 0~, \leqno({\rm H3})
$$
we can give a partial answer to both questions when $1+4\epsilon F'(1) \ge
0$. Indeed, in this case, the Maximum Principle allows us to compare
the solution $w(x,t)$ of \equ(w) with solutions of the linear equations
$$
   \epsilon \tilde w_{tt} + \tilde w_t - 2\epsilon c \tilde w_{xt} \,=\, 
   \tilde w_{xx} + c \tilde w_x + F'(h)\tilde w~, \EQ(first)
$$
and
$$
   \epsilon \tilde w_{tt} + \tilde w_t - 2\epsilon c \tilde w_{xt} \,=\, 
   \tilde w_{xx} + c \tilde w_x + F'(0)\tilde w~. \EQ(second)
$$
We denote by $(\tilde w(t),\tilde w_t(t))$ the solution of \equ(first)
with initial data $(\phi_0,\phi_1)$, and by $(\tilde w^\pm(t),
\tilde w^\pm_t(t))$ the solution of \equ(second) with initial data 
$(\phi^\pm_0,\phi^\pm_1)$, the ``positive'' and ``negative'' parts of 
$(\phi_0,\phi_1)$ defined in \equ(phi+), \equ(phi-). Following the 
arguments in [14, Section~4], we obtain our last stability result:

\CLAIM Theorem(decay) Assume that (H1), (H3) hold and that $1+4\epsilon
F'(1) \ge 0$. Given $c \ge c_*$, $d \in [0,1]$, there exists a constant 
$C_3(c) \ge 1$ such that, for all $K \ge 0$ satisfying
$$
   1 + 4 \epsilon \Lambda_1(K) \,\ge\, 0~,
$$
the following result holds: for any $(\phi_0,\phi_1) \in Z_\epsilon$ 
satisfying \equ(phi0), \equ(phi1) and
$$
   \inf\left(\|(\phi_0,\phi_1)\|_{Z_\epsilon}\,,\, 
   \|(\phi_0^+,\phi_1^+)\|_{Z_\epsilon}\right) \,\le\, {K \over C_3(c)}, 
$$
there exists a unique solution $(w,w_t) \in \CC^0([0,\infty),Z_\epsilon)$ 
of \equ(w) with initial data $(\phi_0,\phi_1)$. Moreover, we have
$$
   -(1-d)h(x) \,\le\, w(x,t) \,\le\, K~, \EQ(wneg)
$$
and
$$
   \tilde w^-(x,t) \,\le\, w(x,t) \,\le\, \tilde w(x,t) \,\le\, 
   \tilde w^+(x,t)~,
$$
for all $x\in \real$, $t \in \real_+$. Finally, if $d > 0$ and 
$1-4\epsilon F'(0) > 0$, one has
$$
   \lim_{t \to +\infty} t^{1/4} \left(\|p_s w(t)\|_{\L^\infty} + 
   \|w(t)\|_{\H^1} + \|w_t(t)\|_{\L^2}\right) \,=\, 0~.
$$

\REMARK If $(\phi^+_0,\phi^+_1) = 0$, {\sl i.e.} if the initial data
are non-positive, then \equ(wneg) with $K=0$ shows that the solution 
$w(x,t)$ remains non-positive for all times. 


\SECTION Asymptotic Expansions in the Critical Case $c=c_*$

In this section, we restrict ourselves to the critical case $c=c_*$,
and we consider perturbations of the travelling wave $h$ in a strict 
subspace of $Z_\epsilon$. Using self-similar variables and energy estimates, 
we are able to compute explicitly the long-time asymptotics of the 
perturbations as $t \to +\infty$. In particular, we recover the results
obtained by Gallay [13] in the parabolic case $\epsilon = 0$. 

Following Kirchg\"assner [25], we consider solutions of \equ(v) 
of the following form
$$
   v(x,t) \,=\, h(x) + h'(x)W\left(x,{t \over 1+\epsilon c_*^2}\right)~,
$$
{\sl i.e.} we set $w(x,t) = h'(x)W(x,t/(1{+}\epsilon c_*^2))$. Then $W$ 
satisfies the equation
$$
   \eta W_{tt} + (1-\nu \gamma(x))W_t - 2\nu W_{xt} 
   \,=\, W_{xx} + \gamma(x) W_x + h'(x)W^2 
   \NN(h(x),h'(x)W)~, \EQ(W)
$$
where 
$$
   \eta \,=\, {\epsilon \over (1+\epsilon c_*^2)^2}~, \quad
   \nu \,=\, {\epsilon c_* \over 1+\epsilon c_*^2}~, \quad
   \gamma(x) \,=\, c_* + 2 {h''(x) \over h'(x)}~, \quad 
   x \in \real~.
$$
In \equ(W) and in the sequel, the second argument of the function $W$ is 
simply denoted by $t$, instead of $t/(1{+}\epsilon c_*^2)$. 

{}From [1] we know that the travelling wave $h$ (with $c=c_*$) satisfies
$$
   h(x) \,=\, \cases{1 + \OO(\e^{\beta x}) & as $x \to -\infty$~, \cr
   (a_1 x+a_2)\e^{-c_* x/2} + \OO(x^2\e^{-c_* x}) & as $x \to 
   +\infty$~,}
$$
where $a_1 > 0$, $a_2 \in \real$, and $\beta = {1 \over 2}(-c_* +
\sqrt{c_*^2-4F'(1)}) > 0$. Similar asymptotic expansions hold
for the derivatives $h'$, $h''$, hence
$$
   \gamma(x) \,=\, \cases{\gamma_- + \OO(\e^{\beta x}) & as $x \to -\infty$~,
   \cr 2/(x{+}x_0) + \OO(x \e^{-c_* x/2}) & as $x \to +\infty$~,}
  \EQ(gamassym)
$$
where $\gamma_- = c_*+2\beta = 2\sqrt{F'(0)-F'(1)}$ and $x_0 = (a_2/a_1-
2/c_*)$. The hypothesis (H1) on $F$ also implies that $\gamma'(x) < 0$ 
for all $x \in \real$. 

To study the long-time behavior of the solution $W$ of \equ(W), we use the 
{\sl scaling variables} or {\sl self-similar variables} defined by
$$
   \xi \,=\, {x \over \sqrt{t{+}t_0}}~,\quad \tau = \log(t{+}t_0)~,
$$
for some $t_0 > 0$. These variables have been widely used to study the 
long time behavior of solutions to parabolic equations, in particular to 
prove convergence to self-similar solutions [4,8,9,12,24]. In [15], it has 
been shown by the authors that these variables are also a powerful tool in 
the framework of damped hyperbolic equations. Following [15], we define the 
rescaled functions $U$ and $V$ by
$$
   U(\xi,\tau) \,=\, \e^{3\tau/2}W(\xi\e^{\tau/2},\e^\tau-t_0)~, \quad 
   V(\xi,\tau) \,=\, \e^{5\tau/2}W_t(\xi\e^{\tau/2},\e^\tau-t_0)~, \EQ(sv)
$$
or equivalently
$$ \eqalign{
   W(x,t) \,&=\, {1 \over (t{+}t_0)^{3/2}} \,U\left({x \over
    \sqrt{t{+}t_0}}\,,\,\log(t{+}t_0)\right)~, \cr
   W_t(x,t) \,&=\, {1 \over (t{+}t_0)^{5/2}} \,V\left({x \over
    \sqrt{t{+}t_0}}\,,\,\log(t{+}t_0)\right)~.} \EQ(vs)
$$
Then $U(\xi,\tau), V(\xi,\tau)$ satisfy the system
$$ \eqalign{
   &U_\tau - {\xi \over 2} U_\xi - {3 \over 2} U \,=\, V~, \cr
   &\eta \e^{-\tau}\bigl(V_\tau - {\xi \over 2}V_\xi - {5 \over 2}V\bigr) + 
   (1-\nu\gamma(\xi \e^{\tau/2}))V -2\nu\e^{-\tau/2}V_\xi \,=\, \cr
   & \qquad \qquad U_{\xi\xi} + \e^{\tau/2} \gamma(\xi\e^{\tau/2}) U_\xi 
   +\e^{-\tau/2} h'(\xi\e^{\tau/2}) U^2 N(\xi,\tau)~,} \EQ(uv)
$$
where $N(\xi,\tau) = \NN(h(\xi\e^{\tau/2}),\e^{-3\tau/2}h'(\xi\e^{\tau/2})U)$.

We now introduce function spaces for the rescaled perturbations $U,V$. 
For $\tau \ge 0$, we denote by $\XX_\tau, \YY_\tau$ the Hilbert spaces of 
measurable functions on $\real$ defined by the norms
$$\eqalign{
  \|V\|_{\YY_\tau}^2 \,&=\, \int_{-\infty}^0 \e^{2\beta \xi\e^{\tau/2}}
   |V(\xi)|^2
   \d \xi + \int_0^{+\infty} (1{+}\xi^6)|V(\xi)|^2\d \xi~, \cr
  \|U\|_{\XX_\tau}^2 \,&=\, \|U\|_{\YY_\tau}^2 + \|U_\xi\|_{\YY_\tau}^2~.}
$$
We denote by $\ZZ_\tau$ the product space $\ZZ_\tau = \XX_\tau \times 
\YY_\tau$ equipped with the standard norm 
$$
   \|(U,V)\|_{\ZZ_\tau}^2 \,=\, \|U\|_{\XX_\tau}^2 + \|V\|_{\YY_\tau}^2~.
$$
As is easily verified, if the functions $(U,V)$ and $(W,W_t)$ are related 
through \equ(sv) or \equ(vs), then $(U(\cdot,\tau),V(\cdot,\tau)) \in 
\ZZ_\tau$ if and only if the actual perturbation $w(x,t) = h'(x)
W(x,t/(1{+}\epsilon c_*^2))$ satisfies
$$
  \int_{-\infty}^0 (w^2 + w_x^2 + w_t^2)(x,t)\d x +
  \int_0^\infty (1+x^4)\e^{c_* x} (w^2 + w_x^2 + w_t^2)(x,t) \d x \,<\, 
  \infty~. \EQ(equiv)
$$
Therefore, the perturbation space considered in this section is slightly
smaller than the space $Z$ defined in \equ(Znorm), due to the factor
$(1+x^4)$ in \equ(equiv). 

Before stating the main result of this section, we explain its 
content in a heuristic way. Taking formally the limit $\tau \to +\infty$ 
in \equ(uv) and using \equ(gamassym), we see that $U$ satisfies the linear 
parabolic equation
$$
   U_\tau \,=\, \LL_\infty U \,\eqdef\, U_{\xi\xi} + \Bigl({\xi \over 2} +
   {2 \over \xi}\Bigr)U_\xi + {3 \over 2} U \quad \hbox{if } \xi > 0~, \quad
   U_\xi \,=\, 0 \quad \hbox{if } \xi \le 0~. \EQ(limiting)
$$
Therefore, it is reasonable to expect that the long-time behavior of
the solutions to \equ(uv) will be determined by the spectral properties 
of the operator $\LL_\infty$ on $\real_+$, with Neumann boundary 
condition at $\xi=0$. Now, as is easily verified, this limiting operator
is just the image under the scaling \equ(vs) of the radially symmetric 
Laplacian operator in three dimensions. Indeed, if $U$ and $W$ 
are related through \equ(vs), the equation $U_\tau = \LL_\infty U$ is 
equivalent to $W_t = W_{xx} + (2/x)W_x$, $x > 0$. 
This crucial observation allows to compute exactly the spectrum of 
$\LL_\infty$ in various function spaces, see [15]. For instance, in 
the space $\L^2(\real_+,(1{+}\xi^6){\rm d}\xi)$, the spectrum of $\LL_\infty$
consists of a simple, isolated eigenvalue at $\lambda=0$, and of 
``continuous'' spectrum filling the half-plane $\{\lambda \in \complex\,|\, 
\Re\lambda \le -1/4\}$. The eigenfunction corresponding to $\lambda = 0$ 
is the gaussian $\e^{-\xi^2/4}$. Therefore, we expect that
the solution $U(\xi,\tau)$ of \equ(uv) converges as $\tau \to +\infty$ to 
$\alpha \phi^*(\xi)$ for some $\alpha \in \real$, where 
$$
   \phi^*(\xi) \,=\, {1 \over \sqrt{4\pi}} \cases{1 & if $\xi < 0$~,\cr
   \e^{-\xi^2/4} & if $\xi \ge 0$~.}
$$
This function is normalized so that $\int_0^\infty \xi^2 \phi^*(\xi)\d \xi =
1$. Since $V = U_\tau - {\xi \over 2}U_\xi -{3 \over 2}U$, we also expect 
that $V(\xi,\tau) \to \alpha \psi^*(\xi)$, where $\psi^* = -{\xi \over 2}
\phi^*_\xi -{3 \over 2}\phi^*$. It is crucial to note that Eq.\equ(limiting)
is independent of $\epsilon$: this explains why the solutions of \equ(W), 
hence of \equ(u), behave for large times in a similar way to those of
the corresponding parabolic equations. 

Our last result shows that the heuristic arguments above are indeed correct:

\CLAIM Theorem(main) Assume that (H1) holds, and let $\epsilon > 0$, 
$c=c_*$. There exist $\tau_0 > 0$ and $\delta_0 > 0$ such 
that, for all $(U_0,V_0) \in \ZZ_{\tau_0}$ satisfying 
$\|(U_0,V_0)\|_{\ZZ_{\tau_0}} \le \delta_0$, the system \equ(uv) has 
a unique solution $(U,V) \in \CC^0([\tau_0,+\infty),\ZZ_\tau)$ with
$(U(\tau_0),V(\tau_0)) = (U_0,V_0)$. In addition, there exists $\alpha^* 
\in \real$ such that
$$
   \|U(\tau)-\alpha^*\phi^*\|^2_{\XX_\tau} + \int_{\tau_0}^\tau 
   \e^{-(\tau-\sigma)/2} \|V(\sigma)-\alpha^*\psi^*\|^2_{\YY_\sigma}\d \sigma
   \,=\, \OO(\tau^2 \e^{-\tau/2})~,
$$
as $\tau \to +\infty$. 

\REMARK We say that $(U,V) \in \CC^0([\tau_0,+\infty),\ZZ_\tau)$ is a 
solution of the system \equ(uv) if there exists a solution $(W,W_t) \in 
\CC^0([0,+\infty),\ZZ_0)$ of \equ(W) such that \equ(sv), \equ(vs) hold,
with $t_0 = \e^{\tau_0}$. 

In terms of the original variables, \clm(main) implies that [16]
$$
   \sup_{x \in \real} \left(1 + {\e^{c_*x/2} \over 1+|x|}\right)
   \left|w(x,t)-{\alpha \over t^{3/2}}\,h'(x)\phi^*\Bigl({x
   \sqrt{1{+}\epsilon c_*^2} \over 
   \sqrt{t}}\Bigr)\right| \,=\, \OO(t^{-7/4}\log t)~,
$$
as $t \to +\infty$, where $\alpha = \alpha^* (1+\epsilon c_*^2)^{3/2}$, 
$w(x,t) = h'(x) W(x,t/(1{+}\epsilon c_*^2))$ and $W$ is given by \equ(vs). 
In the parabolic case $\epsilon = 0$, this
result has been obtained in [13] using slightly different
function spaces. Remarkably enough, the asymptotic profile $\phi^*$ 
is {\sl universal}: it does not depend on the initial data, nor on the 
parameter $\epsilon \ge 0$, nor on the precise form of the non-linearity $F$. 

\REFERENCES

\item{[1]} D.G. Aronson and H.F. Weinberger: 
  {\pap Multidimensional Nonlinear Diffusion Arising in Population Genetics},
  Adv. in Math. {\bf 30} (1978), 33--76.   

\item{[2]} M. Bramson:
  {\bok Convergence of Solutions of the Kolmogorov Equation to Travelling
  Waves}, Memoirs of the AMS {\bf 44}, nb. 285, Providence (1983). 

\item{[3]} J. Bricmont and A. Kupiainen: {\pap Renormalization Group and the
  Ginzburg-Landau Equation}, Commun.\ Math.\ Physics {\bf 150} (1992), 
  193--208. 

\item{[4]} J.  Bricmont and A.  Kupiainen: {\pap Stable Non-Gaussian
  Diffusive Profiles}, Nonlinear Anal., Theory Methods Appl. {\bf 26}
  (1996), 583--593.

\item{[5]} J.  Bricmont, A.  Kupiainen and G.  Lin: {\pap
  Renormalization Group and Asymptotics of Solutions of Nonlinear  
  Parabolic Equations}, Comm.  Pure Appl.  Math.  {\bf 47} (1994),
  893--922.

\item{[6]} S.R. Dunbar and H.G. Othmer: {\pap On a Nonlinear Hyperbolic
  Equation Describing Transmission Lines, Cell Movement, and Branching
  Random Walks}, in {\bok Nonlinear Oscillations in Biology and Chemistry}, 
  H.G. Othmer (Ed.), Lect. Notes in Biomathematics {\bf 66}, Springer (1986). 

\item{[7]} J.-P. Eckmann and C.E. Wayne:
  {\pap The Nonlinear Stability of Front Solutions for Parabolic Partial
  Differential Equations}, Comm. Math. Phys. {\bf 161} (1994), 323--334. 

\item{[8]} M. Escobedo, O. Kavian and H. Matano: {\pap Large Time
  Behavior of Solutions of a Dissipative Semi-linear Heat Equation},
  Comm.  Partial Diff.  Equations {\bf 20} (1995), 1427--1452.

\item{[9]} M. Escobedo and E. Zuazua: {\pap Large-time Behavior
  for Convection Diffusion Equations in $\real^N$}, J.  Funct.  Anal.
  {\bf 100} (1991), 119--161.

\item{[10]} P.C. Fife:
  {\bok Mathematical Aspects of Reacting and Diffusing Systems}, Lect. Notes 
  in Biomathematics {\bf 28}, Springer (1979). 

\item{[11]} R.A. Fisher:
  {\pap The Advance of Advantageous Genes}, Ann. of Eugenics {\bf 7} (1937), 
  355--369. 

\item{[12]} V.A. Galaktionov and J. L. Vazquez:
  {\pap Asymptotic Behaviour of Nonlinear Para\-bolic Equations with Critical 
  Exponents. A Dynamical System Approach}, J. Funct. Anal. {\bf 100} (1991), 
  435--462.

\item{[13]} Th. Gallay:
  {\pap Local Stability of Critical Fronts in Nonlinear Parabolic
  Partial Differential Equations}, Nonlinearity {\bf 7} (1994), 741--764. 

\item{[14]} Th.  Gallay and G.  Raugel: {\pap Stability of Travelling
  Waves for a Damped Hyperbolic Equation}, ZAMP {\bf 48} (1997),
  451--479.

\item{[15]} Th.  Gallay and G.  Raugel: {\pap Scaling Variables and 
  Asymptotic Expansions in Damped Wave Equations}, to appear in 
  J. Diff. Eqns (1998). 

\item{[16]} Th.  Gallay and G.  Raugel: {\pap Scaling Variables
  and Stability of Hyperbolic Fronts}, in preparation.

\item{[17]} S. Goldstein: {\pap On Diffusion by Discontinuous Movements
  and the Telegraph Equation}, Quart. J. Mech. Appl. Math. {\bf 4}
  (1951), 129--156.

\item{[18]} K.P. Hadeler:
  {\pap Hyperbolic Travelling Fronts}, Proc. Edinb. Math. Soc. {\bf 31} 
  (1988), 89--97.   

\item{[19]} K.P. Hadeler:
  {\pap Reaction Telegraph Equations and Random Walk Systems}, 
  in: {\bok Stochastic and spatial structures of dynamical systems},
  S. van Strien, S. Verduyn Lunel (eds.), Royal Acad. of the Netherlands,
  North Holland, Amsterdam (1996).  

\item{[20]} K.P. Hadeler:
  {\pap Reaction transport systems}, in: {\bok Mathematics
  inspired by biology}, V.Capasso, O.Diekmann (eds), CIME Lectures 1997, 
  Florence, Springer Verlag, in print.

\item{[21]} L. Hsiao and T.-P. Liu: {\pap Convergence to Nonlinear
  Diffusion Waves for Solution of a System of Hyperbolic Conservation
  Laws with Damping}, Comm.  Math.  Phys.  {\bf 143} (1992), 599--605.

\item{[22]} M. Kac: {\pap A Stochastic Model Related to the
  Telegrapher's Equation}, Rocky Mountain J. Math. {\bf 4} (1974),
  497--509.

\item{[23]} T. Kapitula:
  {\pap On the Stability of Travelling Waves in Weighted $L^\infty$ 
  Spaces}, J. Diff. Eqns {\bf 112} (1994), 179--215. 

\item{[24]} O. Kavian: {\pap Remarks on the Large Time Behavior of a
  Nonlinear Diffusion Equation}, Ann.  Inst.  Henri Poincar\'e {\bf 4}
  (1987), 423--452.

\item{[25]} K. Kirchg\"assner:
  {\pap On the Nonlinear Dynamics of Travelling Fronts}, J. Diff. Eqns. 
  {\bf 96} (1992), 256--278. 

\item{[26]} A.N. Kolmogorov, I.G. Petrovskii and N.S. Piskunov:
  {\pap Etude de la diffusion avec croissance de la quantit\'e de mati\`ere
  et son application \`a un probl\`eme biologique}, Moscow Univ. Math. Bull.
  {\bf 1} (1937), 1--25. 

\item{[27]} J.D. Murray: {\bok Mathematical Biology} 2nd ed.,
  Biomathematics {\bf 19}, Springer Verlag (1993).

\item{[28]} K. Nishihara: {\pap Convergence Rates to Nonlinear
  Diffusion Waves for Solutions of System of Hyperbolic Conservation
  Laws with Damping}, J.  Differential Equations {\bf 131} (1996),
  171--188.

\item{[29]} M.H. Protter and H.F. Weinberger:
  {\bok Maximum Principles in Partial Differential Equations}, 
  Prentice Hall, Englewood Cliffs N.J. (1967). 

\item{[30]} D.H. Sattinger:
  {\pap On the Stability of Waves of Nonlinear Parabolic Systems}, 
  Adv. Math. {\bf 22} (1976), 312--355. 

\item{[31]} G.I. Taylor: {\pap Diffusion by Continuous Movements},
  Proc. London Math. Soc. {\bf 20} (1920), 196--212.

\item{[32]} A. Volpert, V. Volpert, V. Volpert:
  {\bok Traveling Wave Solutions of Parabolic Systems}, AMS Translations
  of Math. Monographs {\bf 140}, Providence R.I. (1994). 

\item{[33]} E. Zauderer: {\bok Partial Differential Equations of
  Applied Mathematics}, New York, John Wiley (1983).

\end